# Experimental Progress in Ambient-Pressure Superconducting Bilayer Nickelate Films

**Meng Zhang[1,*] and Xi Yan[2,*]**

[1]School of Physics, Zhejiang University, Hangzhou 310027, China
[2]School of Physics, Nankai University, Tianjin 300071, China
[*]E-mail: physmzhang@zju.edu.cn and yanxi@nankai.edu.cn

## Abstract

Bilayer Ruddlesden-Popper nickelates display superconductivity near 80 K under high pressure, establishing a new nickelate platform for studying unconventional high-temperature superconductivity. The recent stabilization of superconducting $RA_3Ni_2O_7$ (RA = rare earth or alkaline earth) films at ambient pressure has changed the experimental landscape: epitaxial strain can reproduce key structural ingredients of the high-pressure phase while making transport, spectroscopy, microscopy, and device-oriented measurements directly accessible. This Review summarizes the experimental progress on ambient-pressure superconducting bilayer nickelate films, with emphasis on synthesis routes, oxygen stoichiometry, substrate-induced strain, normal-state transport, superconducting properties, doping phase diagrams, and momentum-resolved electronic structure. We highlight several issues that remain unsettled, including the reproducibility of phase-pure ultrathin films, the microscopic origin of the two-step superconducting transition, the role of oxygen defects and substrate-derived doping, the position of the Ni $3d_{z^2}$-derived $\gamma$ band, and the pairing symmetry. We close by outlining experimental directions that could establish a more quantitative link among crystal structure, orbital reconstruction, and superconductivity in bilayer nickelate films.

# 1. Introduction

Bilayer Ruddlesden-Popper (RP) nickelate thin films have emerged because epitaxial strain offers a practical route to mimic key structural effects of high pressure without confining the sample in a pressure cell. In bulk $RA_3Ni_2O_7$, superconductivity is induced only when externally applied pressure drives the lattice toward a more symmetric high pressure structure, suppressing octahedral distortions and modifying the Ni-O-Ni bonding geometry [1–3]. Compressive substrates provide an alternative pathway by imposing built-in in-plane lattice compression on epitaxial films, thereby partially reproducing the structural conditions favorable for superconductivity at ambient pressure [4–8]. The immediate advantage of this approach is accessibility: once superconductivity is stabilized at ambient pressure, the films can be examined directly by techniques that are difficult or impossible to implement in a pressure cell, including angle resolved photoemission spectroscopy [9–14], resonant spectroscopies [15,16], scanning tunneling microscopy/spectroscopy [17], mutual inductance [18], high-field magnetotransport [19], and atomic resolution electron microscopy [5,8,20,21].

The significance of the film platform, however, extends beyond accessibility. Unlike pressurized bulk crystals, epitaxial films offer multiple experimental tuning parameters that can be adjusted independently or in combination. Substrate strain can regulate lattice symmetry and octahedral connectivity [6–8]; chemical substitution and oxygen stoichiometry can tune the carrier density and defect landscape [20,22,23]; and heteroepitaxy can engineer interfacial structure, stacking sequence, and dimensionality [9,24]. These attributes make $RA_3Ni_2O_7$ films a uniquely adaptable system for connecting structure, orbital occupancy, electronic correlations, and superconductivity. In this sense, thin films are not simply an ambient-pressure substitute for high-pressure bulk superconductors, but a controllable materials platform for testing the essential ingredients of high-temperature nickelate superconductivity.

Several issues now define the frontier of bilayer nickelate films. On the materials side, the superconducting phase requires a narrow balance among RP phase purity, compressive strain, oxygen stoichiometry, and interfacial continuity [4,5,8,20,21,25,26]. On the physics side, the most active questions concern whether the Ni $3d_{z^2}$-derived $\gamma$ band crosses the Fermi level [9–14], how oxygen defects and substrate induced doping reshape the multiband Fermi surface [10,20,22,23], whether the superconducting gap is nodeless or sign-changing [13,14,17,27–37], and why the maximum superconducting transition temperature ($T_c$) of films still remains below that of pressurized bulk crystals [1,3–5,18]. These questions are strongly coupled, because the same strain fields, defects, and interfaces that stabilize superconductivity may also obscure the intrinsic electronic structure.

This Review focuses on experimental progress in ambient-pressure superconducting bilayer nickelate films; for broader coverage of nickelate superconductivity, the reader is referred to Refs. [38–44] The paper is organized as follows. Section 2 introduces the fundamental properties of nickelates—crystal structure, electronic configuration, and valence state—and traces the research trajectory from infinite-layer to RP-phase materials, with particular emphasis on bulk bilayer nickelate studies, thereby establishing the context for the thin-film discussion. Section 3 describes the synthesis methods for bilayer nickelate films and the principal challenges involved. Section 4 addresses substrate selection and the influence of epitaxial strain on superconducting properties. Sections 5 and 6 examine the normal-state and superconducting-state properties of the films, respectively. Section 7 provides a summary and outlook.

## 2. Discovery of Nickelate Superconductivity

The discovery of nickelate superconductivity represents a recent chapter in the long-standing search for higher superconducting transition temperatures and for material systems beyond the cuprates. Since superconductivity was first observed in mercury in 1911 and high-temperature superconductivity was later established in the cuprates [45] and iron-based superconductors, nickel oxides have attracted sustained attention because Ni-based square-planar or layered structures can, in principle, realize orbital configurations and valence states reminiscent of the $CuO_2$ planes [46]. Early layered nickel compounds, such as LaNiOP [47], LaNiOAs [48], and $LaNiAsO_{1-x}F_x$ [49], exhibited superconductivity only at low temperatures and did not establish a high- $T_c$ nickelate counterpart of the cuprates. The field changed substantially after the discovery of superconductivity in infinite-layer nickelate thin films [50], followed by the observation of liquid-nitrogen-temperature superconductivity in pressurized bilayer RP $La_3Ni_2O_7$ [1] and related RP phases [51–53]. These discoveries have shaped the current development of nickelate superconductivity, tracing a progression from reduced infinite-layer thin films to high-pressure bulk RP compounds, and more recently to ambient-pressure superconducting bilayer nickelate films. The timeline of key discoveries across several representative families of superconducting materials[1–5,9,18,20,24–26,40,50,52–63] is shown in Fig. 1.

### 2.1 Infinite-Layer Nickelates

Infinite-layer nickelates provided the first experimentally established superconducting nickel oxide platform and set the stage for the subsequent exploration of RP nickelates. In 2019, Li et al. reported superconductivity near 15 K in hole-doped $Nd_{0.8}Sr_{0.2}NiO_2$ thin films obtained by topotactic reduction of perovskite $Nd_{0.8}Sr_{0.2}NiO_3$ [50]. This reduced phase contains square-planar $NiO_2$ layers (Fig. 2). The parent compounds are typically $RNiO_2$ (R = La, Pr, Nd, Sm, etc.), doped by alkaline earth substitution such as Sr or Ca; their phase diagrams are summarized in Refs. [54–57,64–66]. The five-layer square-planar nickelate $Nd_6Ni_5O_{12}$ (nominal $3d^{8.8}$), synthesized through an analogous reduction route, also exhibits superconductivity with $T_c$ = 13 K [58]. Infinite-layer nickelates therefore established a new superconducting family based on a nominal $3d^9$ electronic configuration and a square-planar $NiO_2$ framework structurally reminiscent of hole-doped cuprates. Although rare earth 5d states introduce additional hybridization and self-doping channels, the low-energy electronic structure remains strongly shaped by the Ni $3d_{x^2-y^2}$ orbital, allowing selected concepts from cuprate physics to be tested in a distinct nickelate environment.

Recent advancements indicate that the $T_c$ of infinite-layer nickelates can exceed the McMillan limit. For example, epitaxial growth of hole-doped $SmNiO_3$ on $NdGaO_3$ (110) substrates followed by reduction yielded a $T_c$ above 40 K [59]. More recently, freestanding $Nd_{0.85}Sr_{0.15}NiO_2$ films under 91.5 GPa achieved a remarkable $T_c$ of 74.2 K [60]. In this regime, $T_c$ scales almost linearly with applied pressure without apparent saturation, suggesting the potential for even higher transition temperatures. It should be noted that in defect-free infinite-layer nickelate films [65], the normal-state resistivity typically displays a linear temperature dependence, characteristic of strange metal behavior. Finally, it is worth emphasizing that superconductivity in $RNiO_2$ systems has thus far been realized exclusively in thin film form; bulk superconductivity in the infinite-layer phase has not yet been observed and remains an active area of search.

### 2.2 Ruddlesden-Popper Phase Nickelates

The RP structure constitutes another fundamental configuration for nickelates, characterized by perovskite-derived layered geometries with the general formula $RA_{n+1}Ni_nO_{3n+1}$ (where RA represents a rare-earth or alkaline-earth element). This architecture comprises alternating rock salt RAO layers and perovskite $RANiO_3$ blocks stacked along the *c*-axis, where adjacent $NiO_6$ octahedra are corner-sharing ($n \geq 2$). The in-plane Ni–O layers bear a striking resemblance to the Cu–O planes in cuprates, with *n* denoting the number of contiguous Ni–O layers. Depending on *n* and the ionic radii of the RA-site cations, structural distortions occur at ambient pressure, leading to tilted $NiO_6$ octahedra and corrugated Ni–O planes. Typical RP phases include $RA_2NiO_4$ ($n$ = 1), $RA_3Ni_2O_7$ ($n$ = 2), $RA_4Ni_3O_{10}$ ($n$ = 3), and $RANiO_3$ ($n = \infty$, which reduces to an infinite-layer structure upon topotactic reduction). Figure 2 illustrates the evolution of the Ni coordination environment and valence state as a function of *n*.

This Review centers on the $n = 2$ RP phase, $RA_3Ni_2O_7$, commonly referred to as bilayer nickelates. Before the high-pressure discovery, bulk $La_3Ni_2O_7$ was mainly discussed as a correlated metal with density wave tendencies [67–69]. The field changed substantially when Sun et al. reported signatures of superconductivity near 80 K in $La_3Ni_2O_7$ under pressures above 14 GPa (Fig. 3a) [1], placing nickelates beyond the liquid-nitrogen-temperature threshold. Subsequent observations of superconductivity in RP-related systems, including pressurized $La_4Ni_3O_{10}$ [51,52] and mixed-RP $La_5Ni_3O_{11}$ [53], further broadened the nickelate superconducting family. These discoveries also clarified a central limitation: in bulk RP nickelates, superconductivity is tied to high pressure, which restricts direct measurements of the Fermi surface, superconducting gap, and orbital resolved electronic structure.

However, obtaining phase pure bulk samples remains challenging because of competing RP intergrowths. Partial substitution of La by Pr effectively suppresses impurity phases, as demonstrated in polycrystalline $La_2PrNi_2O_7$, which shows cleaner zero resistance behavior, clear diamagnetic response (Fig. 3b), and an enhanced superconducting volume fraction [2]. Further optimization of crystal growth has pushed the onset $T_c$ to 96 K in Sm-doped $La_3Ni_2O_7$ single crystals under pressure, representing the highest $T_c$ reported so far among nickelate superconductors (Fig. 3c) [3].

The superconductivity of bulk $La_3Ni_2O_7$ is closely linked to both oxygen stoichiometry and pressure induced structural changes. Nanoscale diffraction and electron energy loss spectroscopy (EELS) studies reveal strong spatial inhomogeneity in oxygen content and electronic structure, with ligand holes mainly associated with in-plane and inner apical oxygen sites (Fig. 3d) [70]. Because interlayer pairing scenarios rely strongly on Ni $3d_{z^2}$—O $2_{pz}$ coupling, apical oxygen defects can seriously disrupt the electronic structure relevant to superconductivity [37,71–73]. Under pressure, $La_3Ni_2O_7$ undergoes a structural transition from the ambient-pressure *Amam* phase to a higher-symmetry *Fmmm* or *I4/mmm* phase, accompanied by straightening of the Ni–O–Ni bond angle toward 180° (Fig. 3e) [1,74]. Superconductivity emerges robustly near this structural boundary, around 15 GPa, peaks near 80 K, and coexists with anomalous *T*-linear normal-state resistivity in the superconducting pressure regime (Fig. 3f) [40,43,75]. Subsequent studies have shown that, in addition to the canonical bilayer $La_3Ni_2O_7$-$La_3Ni_2O_7$ (2222) stacking, the $La_2NiO_4$-$La_4Ni_3O_{10}$ intergrowth (1313 phase) [76–79], which shares the same $La_3Ni_2O_7$ stoichiometry, also exhibits high-pressure superconductivity.

Despite these advances, the requirement of high pressure severely limits direct spectroscopic access to the superconducting state of bulk $RA_3Ni_2O_7$, leaving its Fermi surface, pairing symmetry, and correlation effects unresolved [27,28,37,71–73]. This limitation naturally motivates the development of epitaxial thin films. In $La_3Ni_2O_7$ and $(La,Pr)_3Ni_2O_7$ films grown on strongly compressive substrates such as $SrLaAlO_4$, superconducting signatures have now been observed at ambient pressure with onset temperatures above 40 K [4,5]. By reproducing key structural effects of pressure while remaining compatible with ARPES, STM/STS, RIXS, and other probes, these films provide the experimental platform for the remainder of this Review.

## 3. Preparation of $RA_3Ni_2O_7$ Thin Films

Currently, single-phase epitaxial $RA_3Ni_2O_7$ films are synthesized in the ultrathin limit (< 10 nm) [4,5]. While their maximum $T_c$ currently trails that of bulk samples—suggesting ample room for optimization via strain engineering and chemical doping—films offer an unparalleled advantage: their superconducting state can be probed directly at ambient pressure. This accessibility is essential for elucidating the superconducting gap symmetry, Fermi surface topology, and electronic correlation strength via advanced spectroscopies. Furthermore, epitaxial growth permits deliberate control over the stacking configuration. For instance, films with $RA_2NiO_4$-$RA_3Ni_2O_7$ (1212) and $RA_3Ni_2O_7$-$RA_4Ni_3O_{10}$ (2323) configurations have been successfully stabilized [9], allowing for systematic studies of dimensionality and interlayer coupling (detailed in Section 6.3). This section outlines the synthesis methodologies and associated challenges.

### 3.1 Synthesis Methodologies

The epitaxial growth of oxide thin films most commonly employs molecular beam epitaxy (MBE) [80] or pulsed laser deposition (PLD, also known as laser MBE) [81] (Figs. 4a–b). PLD offers high deposition rates, good thickness uniformity, and broad material compatibility, making it particularly suited to rapid materials exploration. MBE, by contrast, affords exceptional control over cation stoichiometry and is well-suited to the preparation of atomically precise, high-quality oxide films and heterostructures. Both techniques have been applied successfully to the epitaxial growth of $RA_3Ni_2O_7$ films [4,20,26]. In PLD synthesis, critical parameters include the compositional homogeneity of the solid-state-reacted target, the laser fluence (0.5–1.8 J $cm^{-2}$), and whether *in-situ* ozone annealing is applied during growth. In MBE, the intrinsically low oxidation potential at reduced pressures makes enhancing the oxidation environment during $RA_3Ni_2O_7$ growth a central process-engineering challenge.

Innovative techniques such as Giant Oxidative Atom Layer-by-Layer Epitaxy (GAE) have also been employed to construct specific RP structural motifs (Fig. 4c) [5,82]. Operating on principles similar to PLD, GAE achieves precise control of the stoichiometry of each atomic sublayer by decomposing the RP unit cell into individual atomic layers and fabricating dedicated targets for each. Growth is monitored in real time by reflection high-energy electron diffraction (RHEED), with individual targets sequentially ablated in the appropriate stacking order. For example, $La_3Ni_2O_7$ growth proceeds by cyclically depositing LaO and $NiO_2$ layers in the sequence LaO–$NiO_2$–LaO–$NiO_2$–LaO.

The underlying deposition principle is not novel and has been widely applied in other oxide systems. A key insight distinguishing the successful realization of superconducting $RA_3Ni_2O_7$ films is the necessity of post-deposition ozone oxidation treatment [4]—though some groups have achieved superconductivity by incorporating a strongly oxidizing atmosphere in situ during growth [26]. This treatment is essential for precisely controlling the oxygen stoichiometry required to stabilize the superconducting phase (see Sections 3.2 and 6.2).

### 3.2 Preparation Challenges

The preparation of high-quality, ambient-pressure superconducting $RA_3Ni_2O_7$ films presents several significant challenges. First, as in bulk synthesis, the thermodynamic growth window for phase-pure $RA_3Ni_2O_7$ films is extremely narrow (Fig. 5a) [4], and obtaining a single-phase structure is the primary obstacle. Second, as previously noted, the oxygen stoichiometry is inextricably linked to superconductivity. Early films grown via standard PLD exhibited insulating transport and required rigorous *ex situ* ozone annealing to achieve a superconducting ground state (Fig. 5b) [4]. To address this, they systematically mapped the phase diagram of ozone partial pressure versus annealing duration [25], identifying a narrow optimal pathway that must be followed rigorously to avoid perovskite or higher-RP impurity phases. Shi et al. [26] have also demonstrated the feasibility of fabricating superconducting films via *in-situ* ozone-assisted growth. Substitution of La by Pr ($La_{3-x}Pr_xNi_2O_7$, with $x$ = 0.15 [5], 1 [25], and 2 [26] being the most commonly employed compositions) suppresses competing RP phases, analogously to the strategy used in bulk materials [2]. Lv et al. demonstrated that pre-depositing a $RA_2NiO_4$ buffer layer to optimize interface reconstruction can effectively reduce stacking fault density [21]. Figure 5c shows a scanning transmission electron microscopy (STEM) image of a high-quality $(La, Pr)_3Ni_2O_7$ film that is phase-pure over a large area [5].

Superconducting $RA_3Ni_2O_7$ films are highly susceptible to degradation in air, a phenomenon generally attributed to rapid oxygen diffusion. As shown in Fig. 5d, after 20 days of ambient exposure the onset $T_c$ remains largely unchanged, but the normal-state resistance increases by a factor of two to three and zero resistance is no longer achieved [4]. Storage in a pure oxygen atmosphere retards degradation [5], while below 200 K the physical properties remain essentially stable (Fig. 5e). The current standard protocol is therefore cryogenic storage in liquid nitrogen [25]. To mitigate degradation during atmospheric transfer, a single-unit-cell $SrTiO_3$ epitaxial capping layer [4] or an ~1 nm amorphous $PrBa_2Cu_3O_7$ protective overlayer [26] can be deposited on the $RA_3Ni_2O_7$ film; the latter has been shown to extend ambient-condition stability to more than 100 days. Combined with liquid-nitrogen storage and cryogenic transfer protocols, these measures are expected to substantially facilitate thin-film experimental studies.

Despite progressive improvements to onset and zero-resistance $T_c$ through optimized synthesis and annealing, $RA_3Ni_2O_7$ films frequently exhibit a “two-step transition” in the resistivity near $T_c$ [4,19,25,26,83]. Even when the residual resistance is already very small near ~30 K, a second low-temperature transition persists in the $R(T)$ curve. This feature substantially reduces the zero-resistance $T_c$ and limits the application of spectroscopic probes. This behavior is consistent with granular superconductivity [24,83], likely stemming from a Josephson-coupled network of distinct superconducting domains (Fig. 5f), though direct verification by local probes such as scanning tunneling microscopy is still awaited.

Finally, the thickness of current $RA_3Ni_2O_7$ films typically ranges from 5 to 7 nm. Systematic thickness variation (nominal thicknesses of 3.5–23.5 nm) [26] has shown that the upper critical thickness for single-phase $RA_3Ni_2O_7$ epitaxy on SLAO is approximately 10 nm. Beyond this limit, strain relaxation drives a transformation of the surface region into the trilayer $RA_4Ni_3O_{10}$ phase. This thickness-dependent phase evolution paradoxically renders the $RA_4Ni_3O_{10}$ surface layer a natural protective capping, contributing to prolonged superconducting stability in air.

# 4. Selection of Single-Crystal Substrates

## 4.1 In-Plane Strain and Lattice Symmetry

A critical prerequisite for fabricating superconducting $RA_3Ni_2O_7$ films is the use of a substrate with a relatively small in-plane lattice constant. SLAO (001) single-crystal wafers are tetragonal substrates with excellent lattice matching to cuprates such as $YBa_2Cu_3O_7$ and are widely used for high-temperature superconducting film deposition.

They have become the predominant substrate for $RA_3Ni_2O_7$ films [4,5] and the most established platform for achieving ambient-pressure superconductivity (although superconducting films on $LaAlO_3$ (LAO) substrates have also been reported, their lower in-plane compressive strain yields substantially reduced $T_c$ values [6]; see Section 4.2). In 2024, Cui et al. explored epitaxial $RA_3Ni_2O_7$ growth on a range of substrates spanning a strain range of −4.12% to +3.44%, but all resulting films showed semiconducting or insulating transport behaviour [84]—an outcome attributable to the absence of post-deposition oxygen stoichiometry optimization. Superconducting films were reported by Hwang's group one year later [4]. Unless otherwise noted, this Review discusses $RA_3Ni_2O_7$ films on SLAO substrates.

Figure 6a summarises the lattice-matching relationships between common substrates and $RA_3Ni_2O_7$ [7,85–88]. $SrTiO_3$ (STO) (001), $NdGaO_3$ (NGO) (110), LAO (001), and SLAO (001) are all established substrates for $RA_3Ni_2O_7$ growth. STO and NGO impose tensile strain, while LAO and SLAO impose compressive strain of varying magnitude. To date, superconductivity has been confirmed only in $RA_3Ni_2O_7$ films on LAO (Section 4.2) [6] and SLAO. Achieving superconductivity on LAO is substantially more demanding owing to the weaker in-plane compression, requiring highly optimised growth and annealing conditions. The electrical transport characteristics of early films on all four substrates are shown in Fig. 6b; only the film on SLAO exhibited superconductivity, with an onset $T_c$ of 42 K [8].

High-resolution structural analyses clarify the role of strain: in-plane lattice compression is a universal requirement for $RA_3Ni_2O_7$ superconductivity, whereas out-of-plane (*c*-axis) compression is not. Multi-slice electron ptychography and X-ray diffraction (XRD) [8] confirm that while physical pressure on bulk crystals isotropically compresses the lattice, epitaxial compressive strain on films constricts the in-plane lattice parameters while expanding the *c*-axis (Figs. 6c–f). Importantly, Fig. 6c further reveals that the degree of in-plane compression achieved in films on SLAO has not yet reached the level attained in bulk crystals under high pressure, which provides a structural rationale for the lower $T_c$ of thin films relative to bulk samples. It should also be noted that, analogously to bulk $RA_3Ni_2O_7$ under applied pressure, in-plane compressive strain in thin films promotes a straightening of the Ni–O–Ni bond angle toward 180°, suppressing the octahedral tilting present in the *Amam* phase and thereby enhancing the crystallographic symmetry of the $NiO_6$ network [8]. The critical structural parameter governing superconductivity is therefore this strain-driven symmetrization of the octahedral framework, rather than *c*-axis compression per se.

Given the pivotal role of epitaxial strain, further tuning of in-plane compression is a natural route to enhancing $T_c$. As shown in Fig. 6a, substrates with smaller in-plane lattice constants than SLAO—including $SrPrAlO_4$, $SrNdAlO_4$, $SrSmAlO_4$, $NdCaAlO_4$ (NCAO), and $YAlO_3$ [85–88]—can impose higher compressive strain on $RA_3Ni_2O_7$ films. Using such substrates, or employing them as buffer layers to create a strain gradient, may further increase $T_c$. However, this strategy faces an intrinsic limit: severe in-plane compression exacerbates out-of-plane elongation through the Poisson effect, which may progressively weaken the interlayer coupling mediated by the $d_{z^2}$

orbitals—a channel considered essential for pairing (Section 6.2). Consequently, identifying the optimal balance between in-plane octahedral symmetry enhancement and out-of-plane orbital overlap represents a primary challenge in strain engineering, and may ultimately set an upper bound on the $T_c$ achievable in thin films relative to bulk samples.

### 4.2 Enhancing $T_c$ via Applied Pressure on Films

External pressure can complement epitaxial strain by maintaining in-plane compression while suppressing *c*-axis elongation. Osada et al. applied pressures up to 20 GPa to thin-film samples on various substrates [7] and found that originally non-superconducting films progressively suppressed density-wave order and developed superconductivity with increasing pressure; films on LAO substrates achieved $T_c$ values up to 60 K (Figs. 7a–c). This study confirmed a positive correlation between $T_c$ and the *c/a* ratio and demonstrates that strain engineering—through modulation of the crystal-field environment and band structure—is a viable route for optimising superconductivity in $RA_3Ni_2O_7$ films. Similarly, $(La,Pr)_3Ni_2O_7/(La,Sr)_3Ni_2O_7$ heterostructures epitaxially grown on SLAO exhibit a marked increase in $T_c$ upon pressure application (Fig. 7d) [24].

Recent optimizations have also yielded ambient-pressure superconductivity (~10 K) in pristine $RA_3Ni_2O_7$ films grown directly on LAO substrates (Fig. 7e) [6]. Because LAO imparts a milder compressive strain (−1.2%) compared to SLAO (−2.0%), early attempts failed to reach the superconducting phase boundary. This recent success lowers the empirical strain threshold required for superconductivity, providing an expanded parameter space to investigate the evolution of the superconducting order parameter across the strain-temperature phase diagram.

## 5. Normal-State Properties of $RA_3Ni_2O_7$ Films

### 5.1 From Fermi Liquid to Non-Fermi Liquid Behavior

The normal-state electrical transport of $RA_3Ni_2O_7$ superconducting films is highly sensitive to the chosen substrate and the specific oxygen stoichiometry. Unconventional superconductivity often emerges from a strongly correlated normal state, typically manifesting as either a highly renormalized Fermi liquid or an anomalous "strange metal" characterized by a strictly $T$-linear resistivity [61]. Magnetotransport studies [19] conducted under pulsed magnetic fields up to 64 T have revealed conventional Fermi liquid behavior in certain $RA_3Ni_2O_7$ films (Figs. 8a–b). In these instances, the resistivity and Hall angle follow a $T^2$ dependence, and the transverse magnetoresistance adheres to Kohler's rule ($H^2$ scaling). These measurements yielded a quasiparticle effective mass of approximately 10 $m_e$, a value consistent with strong electronic correlations (Section 6.3).

Conversely, when $RA_3Ni_2O_7$ films are epitaxially grown on LAO substrates, the normal-state resistivity exhibits a $T^{5/3}$ dependence (Fig. 8c) [6]. Furthermore, systematic variation of the oxygen content [18] demonstrates that as oxygen deficiency increases, the temperature exponent of the resistivity transitions gradually from 2 to 1. In severely oxygen-deficient films [22], the high-temperature resistivity is $T$-linear, eventually crossing over to an ln $T$ dependence at low temperatures. This contrasts markedly with the ubiquitous non-Fermi-liquid behavior observed in bulk $RA_3Ni_2O_7$ crystals under pressure [75], where $T$-linear resistivity is a robust feature of the normal state across the superconducting phase. $T$-linear resistivity is extensively documented in the strange-metal regime of cuprate superconductors and is widely associated with non-Fermi-liquid scattering beyond the quasiparticle paradigm. Taken together, these experimental observations establish that, in sharp contrast to bulk materials, the epitaxial strain environment and oxygen stoichiometry act as dominant tuning parameters governing the normal-state behavior of thin films, offering a versatile handle for exploring the crossover between Fermi-liquid and strange-metal regimes.

### 5.2 Hall Coefficient

While the Hall coefficient ($R_H$) of $RA_3Ni_2O_7$ films varies significantly depending on synthesis conditions and oxygen content, a temperature-dependent sign reversal is frequently observed, providing direct evidence of the system's multiband nature. Figure 8d illustrates the temperature dependence of $R_H$ for films synthesized by different groups [4,5,20,24–26]. Beyond variations in the absolute magnitude, there are distinct differences in the functional form of $R_H$ ($T$) and the dominant carrier type across samples—a reflection of the strong sensitivity of the multiband Fermi surface to preparation conditions. Systematic vacuum annealing experiments performed on a single sample (Figs. 8e–f) explicitly link this behavior to oxygen stoichiometry [22]. In the over-oxidized state, $R_H$ is negative and varies only marginally with temperature. However, in the under-oxidized state, $R_H$ is positive at low temperatures and transitions to negative upon heating, unambiguously confirming the coexistence of competing electron and hole pockets. The observed evolution of the temperature exponent in resistivity (Section 5.1) and the sign reversal in $R_H$ can thus be understood within a unified multiband picture, in which the relative contribution of electron-like and hole-like bands to transport shifts systematically with oxygen stoichiometry and strain.

It is noteworthy that, in bulk $RA_3Ni_2O_7$, a variety of electronic ordered states have been identified in the normal state through RIXS, μSR, NMR, and electrical transport measurements [89–94]. These include density-wave order—encompassing both spin-density-wave and charge-density-wave components—as well as a possible double-stripe spin structure (Fig. 3f). Quantitative consensus on key physical parameters, such as the spatial modulation wavevector of the density-wave order and the magnitude of the magnetic superexchange interaction, has not yet been reached. Strikingly, in $RA_3Ni_2O_7$ thin films, analogous density-wave ordering has not been systematically observed [95], likely owing to the modified electronic environment imposed by the substrate—whether through epitaxial strain, interfacial charge transfer, or suppressed structural symmetry-breaking—though a definitive mechanistic understanding is still lacking.

## 6. Superconducting Properties of $RA_3Ni_2O_7$ Thin Films

### 6.1 Highest Reported $T_c$ and Transport Characterization

To date, the maximum $T_c$ achieved in $RA_3Ni_2O_7$ thin films is 63 K (Fig. 9a) [18], which remains lower than the peak $T_c$ of pressurized bulk samples. Currently, mutual inductance measurement is the sole reported technique capable of resolving the superconducting diamagnetic signal in these films (Fig. 9b). Because the films are ultrathin and possess a minuscule superconducting volume fraction, the diamagnetic response is easily obscured by the large magnetic background of the single-crystal substrate; quantitative extraction of the superconducting volume fraction therefore remains an outstanding experimental challenge. The inability of conventional SQUID magnetometry to overcome this background signal underscores the need for alternative local probes of the superconducting condensate.

The persistent gap between the thin-film and bulk $T_c$ values likely reflects the superposition of several distinct mechanisms. In addition to the insufficient degree of in-plane lattice compression relative to the high-pressure bulk state discussed above, *c*-axis elongation may weaken the interlayer Ni $3d_{z^2}$ orbital coupling that is thought to be central to pairing. Interface scattering and structural disorder can reduce the superconducting coherence length, while the limited film thickness may preclude the formation of a fully three-dimensional superconducting condensate. Granular superconductivity (Section 3.2) further suppresses the zero-resistance $T_c$. Disentangling the relative contributions of these mechanisms remains an important priority for future work.

Through optimized growth and annealing protocols, the critical current density ($J_c$) of these films has reached approximately 10 kA cm$^{-2}$ at 2 K (Fig. 9c) [25]. The films also exhibit exceptionally high upper critical fields [19]: approximately 43 T perpendicular to the *ab*-plane and up to 106 T parallel to it (Fig. 9d). The perpendicular value was determined from direct high-field measurements and is consistent with earlier Ginzburg–Landau estimates; the parallel value, being inaccessible to direct measurement within current pulsed-field capabilities, was obtained by extrapolation and should be regarded accordingly. Ginzburg–Landau analysis yields an out-of-plane coherence length and effective superconducting thickness both in the range of 2–4 nm. Furthermore, the angle-dependent magnetotransport conforms excellently to the two-dimensional Tinkham model, underscoring the strongly two-dimensional character of the superconducting state.

### 6.2 Doping Phase Diagram

Carrier doping is an indispensable tool for tuning the electronic ground states of quantum materials. In high- $T_c$ cuprates, hole doping suppresses the parent antiferromagnetic Mott insulating state and engenders a rich phase diagram featuring pseudogap, strange metal, and Fermi liquid regimes. Given the multiband architecture of the $La_3Ni_2O_7$ system—comprising hybridized La 5*d* and Ni 3*d* orbitals—mapping its phase diagram is paramount. Theoretical models propose that while compressive strain shifts the $d_{z^2}$-derived γ band away from the Fermi level, hole doping can pull it back [29,96,97], making the experimental realization of a doping phase diagram highly significant.

Chemical substitution at the La site with aliovalent cations (e.g., Sr, Ca) is the most direct method of hole doping. Recent studies [20] reveal that $T_c$ remains relatively stable for Sr substitution levels up to $x$ = 0.21, beyond which superconductivity is rapidly suppressed (Fig. 10a). The optimally doped Sr-substituted film reached a $T_c$ of 42 K. STEM analyses indicate that, unlike in bulk samples where vacancies predominantly occupy the inner apical oxygen sites, oxygen vacancies in the films preferentially localize within the $NiO_2$ planes—a likely consequence of the imposed in-plane compressive strain [20].

Due to the extreme sensitivity of $RA_3Ni_2O_7$ films to their oxygen sublattice, oxygen non-stoichiometry serves as a potent alternative doping mechanism. Independent studies [22,23] have successfully mapped a "dome-shaped" phase diagram spanning the under-oxidized and over-oxidized regimes (Figs. 10b–c). Moving away from the optimal superconducting state by increasing the oxygen content systematically suppresses $T_c$ and drives the system toward a conventional metallic phase (Fig. 10b). This semi-dome structure can be understood in terms of the competing effects of interstitial oxygen and oxygen vacancies: a crossover between carrier doping and impurity scattering as the dominant mechanism. Conversely, driving the system into an under-oxidized state induces a transition from the optimal superconductor to a robust insulator (Fig. 10c).

### 6.3 Fermi Surface and Pairing Symmetry

#### 6.3.1 Background

It is instructive to contrast the electronic structure of $RA_3Ni_2O_7$ with that of the cuprates. Cuprate superconductors are based on single or multiple $CuO_2$ planes within a perovskite framework. The Cu $3d^9$ configuration, together with crystal-field splitting and the Jahn–Teller effect, lifts the $d$-orbital degeneracy such that the Fermi level is dominated by the Cu $3d_{x^2-y^2}$ orbital and its hybridized O $2p_{x/y}$ counterparts [61], with the $3d_{z^2}$ orbital lying ~1-2 eV below the Fermi energy [98]. In $RA_3Ni_2O_7$, by contrast, the nominal Ni $3d^{7.5}$ configuration and the bilayer $NiO_2$ structural unit lead to a fundamentally different orbital landscape. The two $NiO_2$ planes are coupled vertically by $\sigma$ bonds formed between the interlayer Ni $3d_{z^2}$ and O $2p_z$ orbitals, and the crystal-field splitting between the two $e_g$ orbitals ($3d_{z^2}$ and $3d_{x^2-y^2}$) is considerably smaller than in the cuprates, resulting in both orbitals carrying substantial spectral weight near the Fermi level [1,99].

Early theoretical treatments [1,99] of the high-pressure superconducting state considered that the $\sigma$-bonding band—formed by hybridization between Ni $3d_{z^2}$ and apical O $2p_z$ at ambient pressure—lies below the Fermi level, but moves upward and crosses the Fermi energy under applied pressure, thereby metallizing and contributing to superconductivity. The $3d_{x^2-y^2}$ orbital, being more itinerant, crosses the Fermi level at both ambient and high pressure, contributing the $\alpha$ and $\beta$ Fermi pockets. Under high pressure, the $3d_{z^2}$ orbital crosses the Fermi energy, forming the $\gamma$ pocket near the M ($\pi$, $\pi$) point. Ambient-pressure ARPES on bulk crystals [100] confirms that the $\gamma$ band lies approximately 50 meV below the Fermi level; direct measurements under pressure remain technically limited.

Regarding the driving force for pairing under high pressure, the principal theoretical proposals [72,101–104] include: interlayer superexchange interaction via the $3d_{z^2}$ orbitals, transmitted to the $3d_{x^2-y^2}$ electrons through Hund coupling between the $e_g$ orbitals or through the enhanced metallicity of the $\sigma$-bonding band; and intra- or interlayer correlation effects within the $3d_{x^2-y^2}$ manifold. These mechanisms are not mutually exclusive; their relative contributions may vary under different physical conditions.

In terms of pairing symmetry, it is useful to recall the contrasting paradigms established in simpler systems. Conventional *s*-wave superconductors exhibit a momentum-isotropic gap with no nodes on the Fermi surface. In cuprates, pairing is driven by nearest-neighbor antiferromagnetic correlations within the approximately square $CuO_2$-derived Fermi sheet, giving rise to a nodal *d*-wave order parameter [105]. In multiband systems where the effective electron–electron attraction is spatially localized—as in interlayer or inter-orbital channels—the pairing symmetry tends instead toward an $s\pm$ state in which the gap function changes sign between different Fermi pockets [106,107].

In $RA_3Ni_2O_7$, both the $3d_{x^2-y^2}$ and $3d_{z^2}$ degrees of freedom appear near the Fermi level. The interlayer superexchange via $3d_{z^2}$, intra-atomic Hund coupling between the two $e_g$ orbitals on the same Ni site, and intra-layer nearest-neighbor correlations within the $3d_{x^2-y^2}$ manifold all contribute to the effective pairing interaction. In general, spatially localized interlayer and inter-orbital exchange interactions favor $s\pm$ pairing, while intra-layer nearest-neighbor correlations in the $3d_{x^2-y^2}$ channel may promote *d*-wave pairing analogous to the cuprate mechanism. Currently, theoretical models explore both scenarios [27,29–35].

### 6.3.2 $RA_3Ni_2O_7$ thin films

The precise electronic structure and pairing symmetry of $RA_3Ni_2O_7$ are currently the subject of intense scrutiny. Because bulk superconductivity is exclusively pressure-induced, ambient-pressure superconducting thin films serve as an indispensable platform for spectroscopic interrogation. Structurally analogous to the critical $CuO_2$ planes in cuprates, the fundamental units of $RA_3Ni_2O_7$ are quasi-two-dimensional $NiO_2$ bilayers; the crystal field splitting between the two transition-metal $e_g$ orbitals ($d_{x^2-y^2}$ and $d_{z^2}$) is minute, resulting in both orbitals holding substantial spectral weight near the Fermi level. Consequently, a fierce theoretical debate has emerged regarding whether the $d_{z^2}$-derived $\gamma$ band actively participates in superconducting pairing [27–29,34,37,73].

In $RA_3Ni_2O_7$ thin films on compressive substrates, the observation of the $\gamma$ band at the Fermi surface by some groups [10] appears to conflict with the simple $\sigma$-band metallization picture [1], since $c$-axis elongation is expected to weaken vertical Ni $3d_{z^2}$–O $2p_z$ hybridization and thereby lower the $d_{z^2}$ orbital energy. Li et al. proposed that unintentional Sr diffusion from the SLAO substrate introduces hole doping, which lowers the chemical potential and restores the $\gamma$ band to the Fermi surface [10]. This interpretation, if correct, would have far-reaching implications: it suggests that the electronic structure measured in virtually all SLAO-based thin-film experiments may be systematically influenced by extrinsic substrate-derived doping, an effect that deserves careful experimental scrutiny—for example, through the use of substrates free of Sr. Consistent with the importance of the $\gamma$ band, Nie et al. synthesized films with four distinct stacking configurations on SLAO substrates and correlated their transport properties with measured Fermi-surface topologies, identifying the $\gamma$ band contribution as vital for superconductivity (Fig. 11) [9]. This effort stabilized two new superconducting architectures (1212 and 2323 phases) and revealed a notable correlation: the 1313 phase, which superconducts in bulk but not in thin-film form, is also associated with the absence of the $\gamma$ band at the Fermi surface in the thin-film geometry. Complementary RIXS and XAS measurements further confirm that interlayer coupling mediated by the $d_{z^2}$ orbitals is integral to the superconducting state [15,16].

Conversely, Wang et al., using identical SLAO substrates, observed the $\gamma$ band residing ~70 meV below the Fermi level [11]; consistent behavior has been reported in independent ARPES studies [12]. The origin of this discrepancy—which may reflect differences in sample preparation, post-annealing conditions, oxygen stoichiometry, or the degree of substrate-induced Sr doping—has not been established. Resolving this fundamental inconsistency is one of the most pressing outstanding problems in the field, as the presence or absence of the $\gamma$ band at the Fermi level directly determines which class of theoretical pairing model is applicable.

While density functional theories predominantly predict either $s\pm$ or $d$-wave symmetry [27–37], scanning tunneling spectroscopy (STS) provides compelling experimental support for an $s\pm$ pairing mechanism. High-resolution scanning tunneling microscopy (STM) data [17] reveal a distinct double-gap structure with magnitudes of 19 meV and 7 meV (Fig. 12a). Modeling these spectra suggests that the dominant pairing channel arises from an anisotropic $s$-wave gap (Figs. 12b–d). The concurrent observation of a strong bosonic mode near 30 meV provides further evidence for a sign-reversing gap. Furthermore, separate ARPES experiments [13,14] have also observed a nodeless superconducting gap. The results resolved a superconducting gap of $\Delta \sim 18$ meV within the $d_{z^2}$-derived band, yielding a coupling ratio of $2\Delta/k_BT_c \sim 8$ that vastly exceeds the BCS weak-coupling limit. These measurements also uncovered an orbital-dependent dimensional effect (Fig. 12e): a highly two-dimensional $d_{x^2-y^2}$ band coexisting with a strongly $k_z$-dispersive $\gamma$ band.

The degree of electron correlation in $RA_3Ni_2O_7$ has itself been debated—specifically whether the system should be classified as moderately or strongly correlated (Fig. 12f) [19]. Recent evidence [70,108] suggests that the correlation character of $RA_3Ni_2O_7$ more closely resembles that of charge-transfer-type cuprates, which may underlie its relatively high $T_c$. Recent spectroscopic measurements on bulk $RA_3Ni_2O_7$ further support a strongly correlated electron picture [70]. Additionally, the two $e_g$ orbitals appear to carry distinct correlation strengths: the $3d_{z^2}$ orbital exhibits stronger renormalization, with a mass-enhancement factor estimated at 5–8 [100], compared to the more itinerant $3d_{x^2-y^2}$ orbital [109–111]. Whether a pseudogap—analogous to the celebrated normal-state feature in underdoped cuprates—exists in $RA_3Ni_2O_7$ thin films remains an open question that has not yet been addressed experimentally, and its investigation could provide a crucial test of the cuprate analogy.

In summary, key questions regarding the band structure and pairing symmetry of $RA_3Ni_2O_7$ thin films remain unresolved. Whether the $\gamma$ pocket appears at the Fermi level, and whether the superconducting gap is nodeless, are both still open issues. Extrinsic factors—including Sr diffusion from the SLAO substrate [10], deviations in oxygen stoichiometry, and intrinsic interface charge transfer—may all influence the measured electronic structure and should be carefully evaluated when interpreting spectroscopic results. For a systematic side-by-side comparison of the electronic properties of $RA_3Ni_2O_7$ and the cuprates, the reader is referred to the summary table in Ref. [41].

## 7. Summary and Outlook

The discovery of nickelate superconductivity is arguably one of the most transformative events in recent condensed matter physics, revitalizing the global pursuit of unconventional high-temperature superconductors. Building upon the foundational physics of infinite-layer and bulk RP-phase nickelates, this Review has synthesized the current experimental landscape of $RA_3Ni_2O_7$ epitaxial thin films. Moving forward, the field must navigate several critical challenges to unlock the full potential of these materials.

**Material synthesis and reproducibility:** Synthesizing high-quality $RA_3Ni_2O_7$ films remains an extraordinary metallurgical challenge, achieved by only a fraction of groups worldwide. Variations in sample quality currently represent the largest barrier to systematic study. Developing highly reproducible, thermodynamically controlled synthesis pathways—coupled with precise oxygen defect engineering and long-term ambient stabilization techniques—is an absolute prerequisite for future advancements.

**Further enhancement of superconducting properties:** Although using substrates with smaller lattice constants to increase in-plane compression may be self-limiting owing to the disruption of interlayer coupling through *c*-axis elongation, alternative strategies—such as doping with large-ionic-radius elements (e.g. K, Ba) to increase the in-plane lattice constant of the film—may offer new routes to higher $T_c$. Furthermore, establishing robust, standardized methodologies for quantifying the superconducting volume fraction and locally mapping the oxygen stoichiometry is essential.

**Elucidating the pairing mechanism:** A unified consensus on the Fermi surface topology and pairing symmetry is required to decode the nickelate superconducting mechanism. In particular, resolving the discrepancy in $\gamma$-band position between different groups, understanding whether a pseudogap phase exists in the normal state, and definitively benchmarking these features against the cuprate paradigm will rely critically on higher-quality, phase-pure thin films and on experimental approaches that explicitly control or characterize substrate-induced interfacial effects.

**Artificial superlattices and device applications.** Artificial superlattices composed of distinct nickelate family members—for example, $RA_3Ni_2O_7/RA_2NiO_4$ or $RA_3Ni_2O_7/RANiO_2$ heterostructures—offer a versatile platform for engineering the interfacial electronic structure. The proximity between layers with distinct orbital characters and carrier densities may promote interface-enhanced superconductivity, orbital reconstruction, and superconducting proximity effects. The fabrication and characterization of Josephson junctions and related superconducting quantum devices based on $RA_3Ni_2O_7$ thin films represent a further avenue with potential implications for quantum information technology.

**Exploration of novel nickelate superconductors:** The recent discovery of high-pressure superconductivity in the $n$ = 3 RP phases, such as $La_4Ni_3O_{10}$ and $Pr_4Ni_3O_{10}$, has significantly broadened the nickelate family. To date, ambient-pressure superconductivity has not been stabilized in $RA_4Ni_3O_{10}$ thin films. Because the bulk $RA_4Ni_3O_{10}$ phase requires even greater physical pressure to superconduct than the $RA_3Ni_2O_7$ phase, utilizing substrates with substantially smaller lattice parameters than SLAO (such as NCAO) may provide the extreme compressive strain necessary to realize ambient-pressure superconductivity in these trilayer systems.

During the preparation of this Review, we noted a concurrent Review [44] by Zhang et al. covering ambient-pressure superconducting films in the $RA_3Ni_2O_7$ phase, further reflecting the rapid and dynamic pace of development in this field.

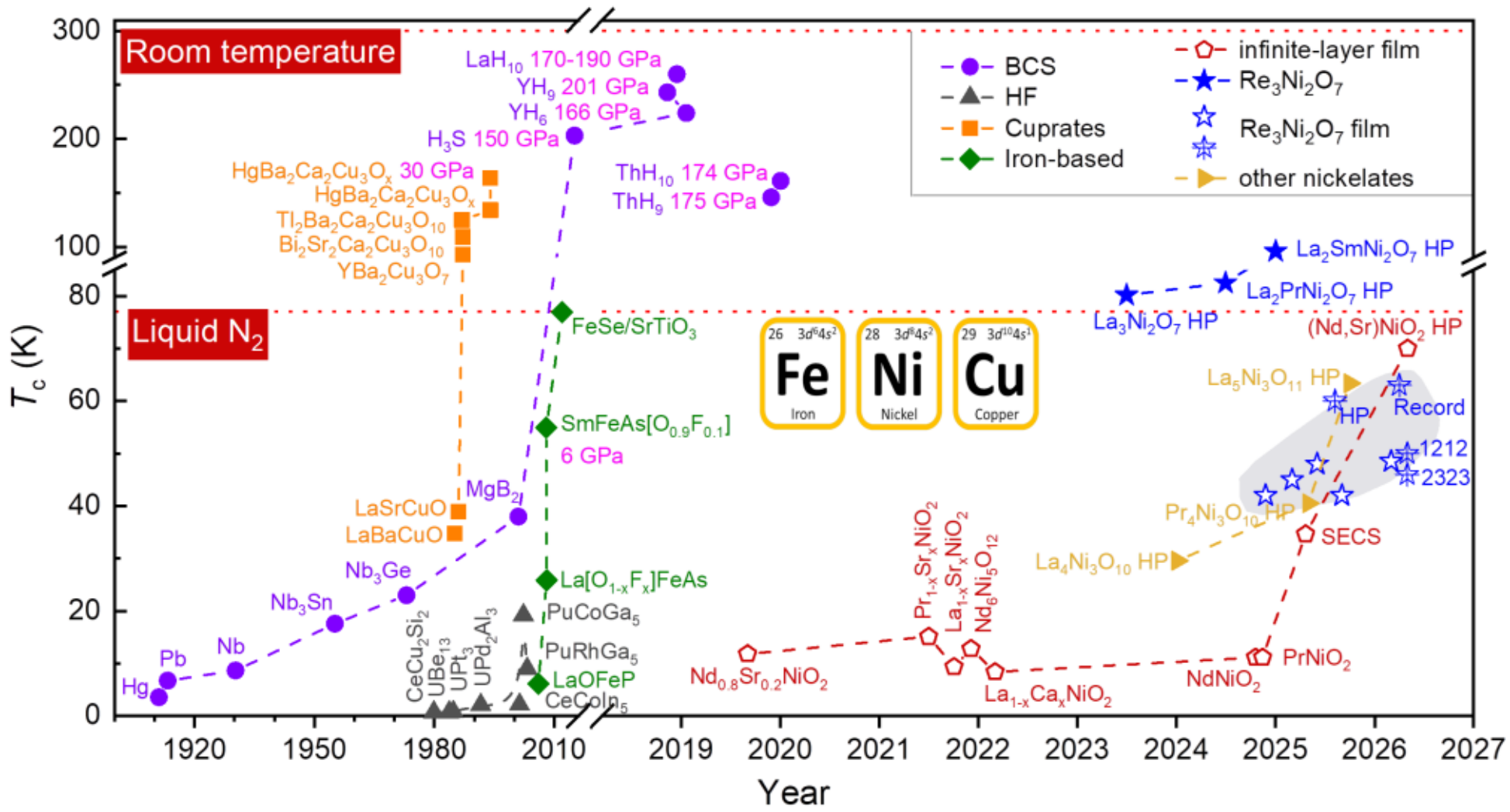

**Figure 1.** Timeline of superconductivity discoveries across material families. The superconducting transition temperature, $T_c$, is plotted against the year of discovery for conventional (BCS), heavy-fermion (HF), cuprate, iron-based, and the recently discovered nickelate superconductors. The nickelate family includes infinite-layer (IL) compounds, bilayer RP phases ($RA_3Ni_2O_7$), and other structural variants. Open symbols denote thin-film samples. Open stars with crosses represent specific $RA_3Ni_2O_7$ thin-film formulations (as labeled). HP: high pressure. Record: the highest achieved $T_c$ in $RA_3Ni_2O_7$ thin films. Designations 1212 and 2323 refer to alternating $RA_2NiO_4$-$RA_3Ni_2O_7$ and $RA_3Ni_2O_7$-$RA_4Ni_3O_{10}$ intergrowth structures, respectively. The shaded region highlights the central focus of this review: ambient-pressure superconducting $RA_3Ni_2O_7$ thin films. Inset: Core elements and electron configurations of the three leading high-temperature superconducting families. The horizontal red dotted lines indicate the boiling point of liquid nitrogen (77 K) and room temperature (300 K).

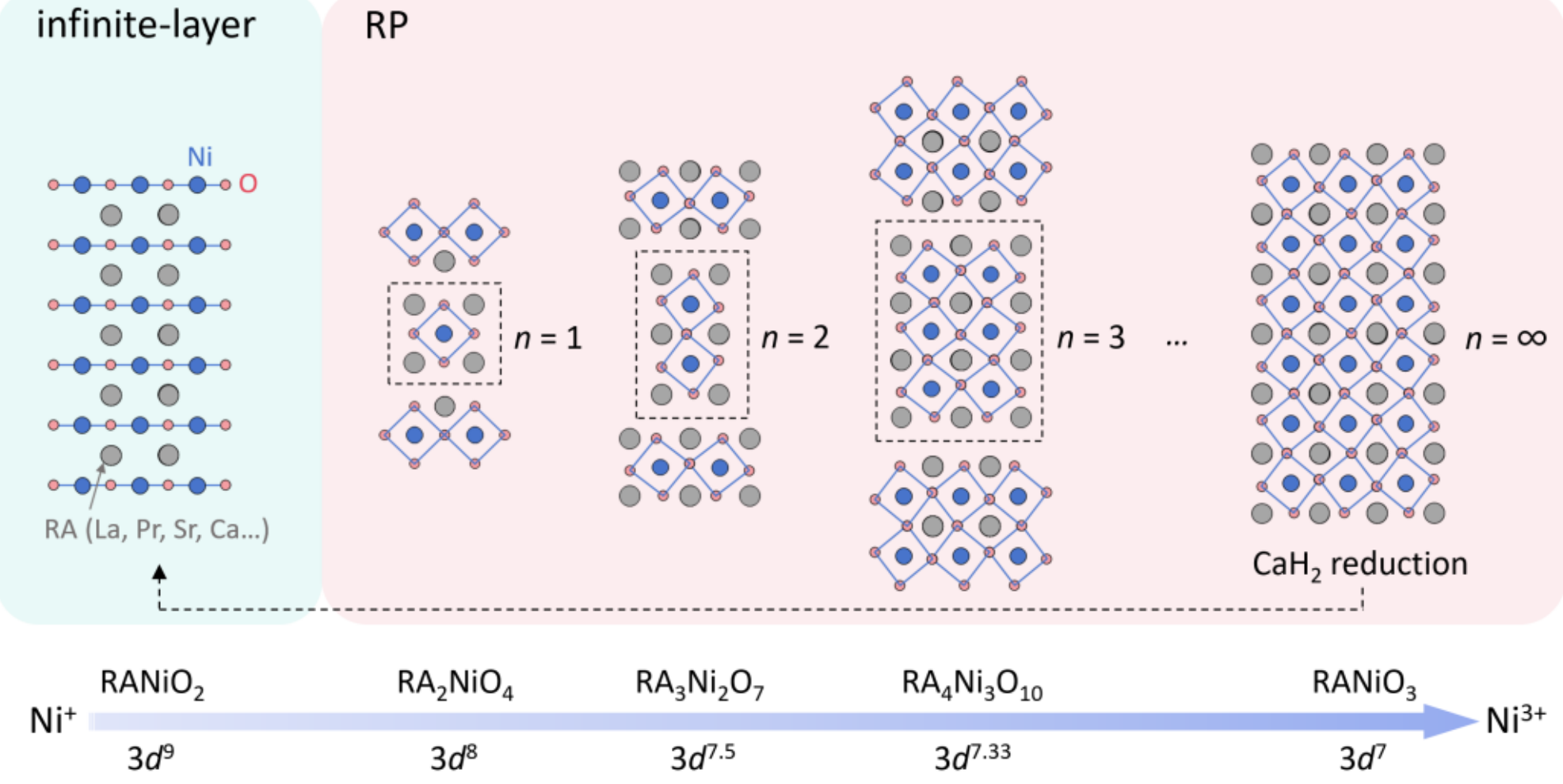

**Figure 2.** Crystal structure, electron configuration, and Ni valence states in nickelate families. The first nickelate superconductor reported was hole-doped $RANiO_2$ (RA = rare earth or alkaline earth),

which adopts an infinite-layer structure [50]. The first nickelate superconductor with a transition temperature above the liquid-nitrogen threshold belongs to the Ruddlesden–Popper (RP) series $RA_{n+1}Ni_nO_{3n+1}$ at $n = 2$ [1], where $n$ denotes the number of $NiO_2$ layers; $RANiO_3$ corresponds to the $n = \infty$ limit. $RANiO_2$ is obtained from $RANiO_3$ via topotactic chemical reduction. In nickelates, the formal Ni valence ranges from +1 to +3, with $Ni^{2+}$ being the most prevalent oxidation state.

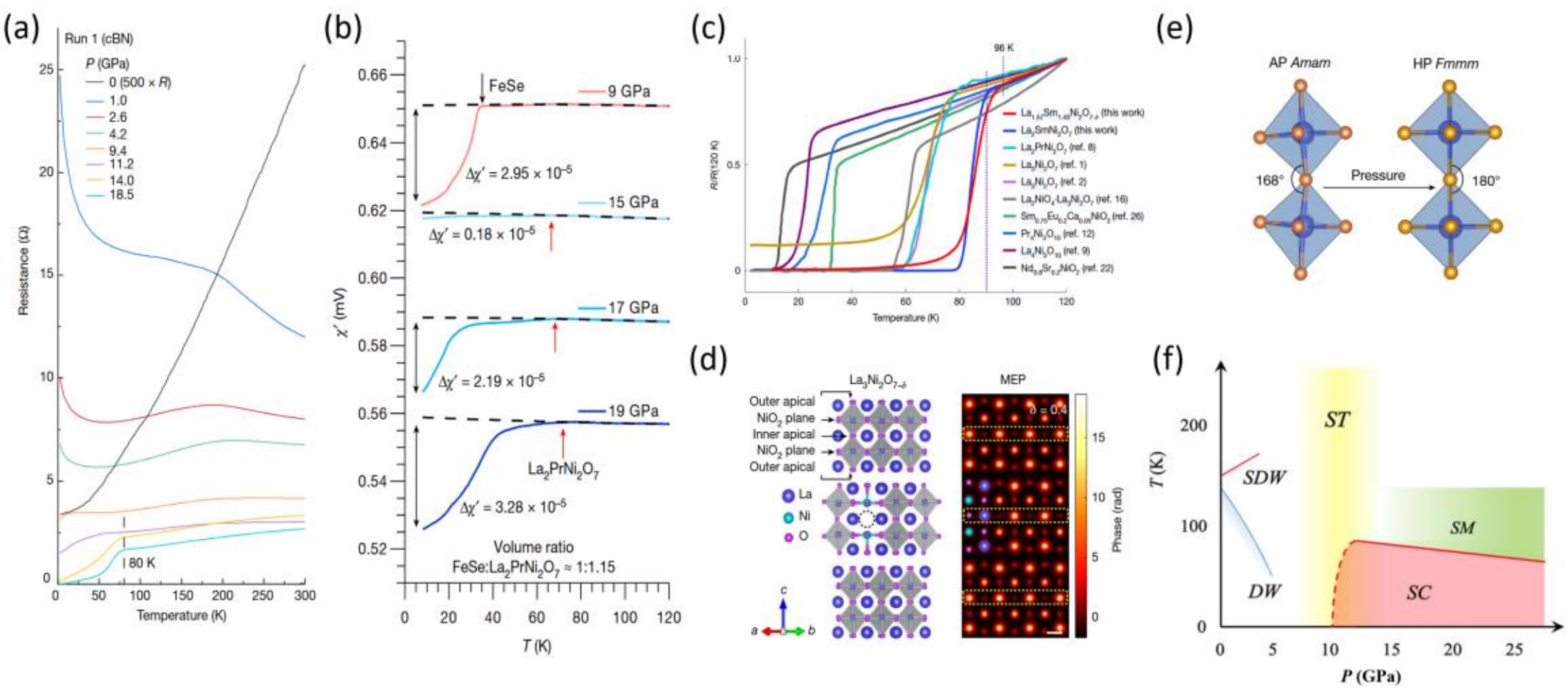

**Figure 3.** Pressure-induced superconductivity in bulk $RA_3Ni_2O_7$. (a) First observation of superconductivity near 80 K in $La_3Ni_2O_7$ single crystals under applied pressure. (b) The superconducting diamagnetic response in $La_2PrNi_2O_7$ polycrystals under high pressure. (c) Maximum $T_c$ values reported for various bilayer nickelates; a $La_{1.57}Sm_{1.43}Ni_2O_{7-\delta}$ single crystal exhibits an onset $T_c$ as high as 96 K under pressure. (d) Schematic of the three distinct oxygen crystallographic sites (outer apical, planar, and inner apical) in the $La_3Ni_2O_7$ structure, illustrating potential oxygen vacancy locations. (e) Pressure-driven structural transition: Ni-O-Ni bond angle between adjacent $NiO_6$ octahedra straightens from 168° (space group *Amam*) at ambient pressure to 180° (*Fmmm*) under high pressure. (f) Schematic temperature-pressure (*T-P*) phase diagram [43] for $La_3Ni_2O_7$, where SDW, DW, ST, SM, and SC denote spin-density wave, density wave, structural transition, strange metal, and superconductivity, respectively. (a-e) Adapted from [1–3,70], with permission from Springer Nature.

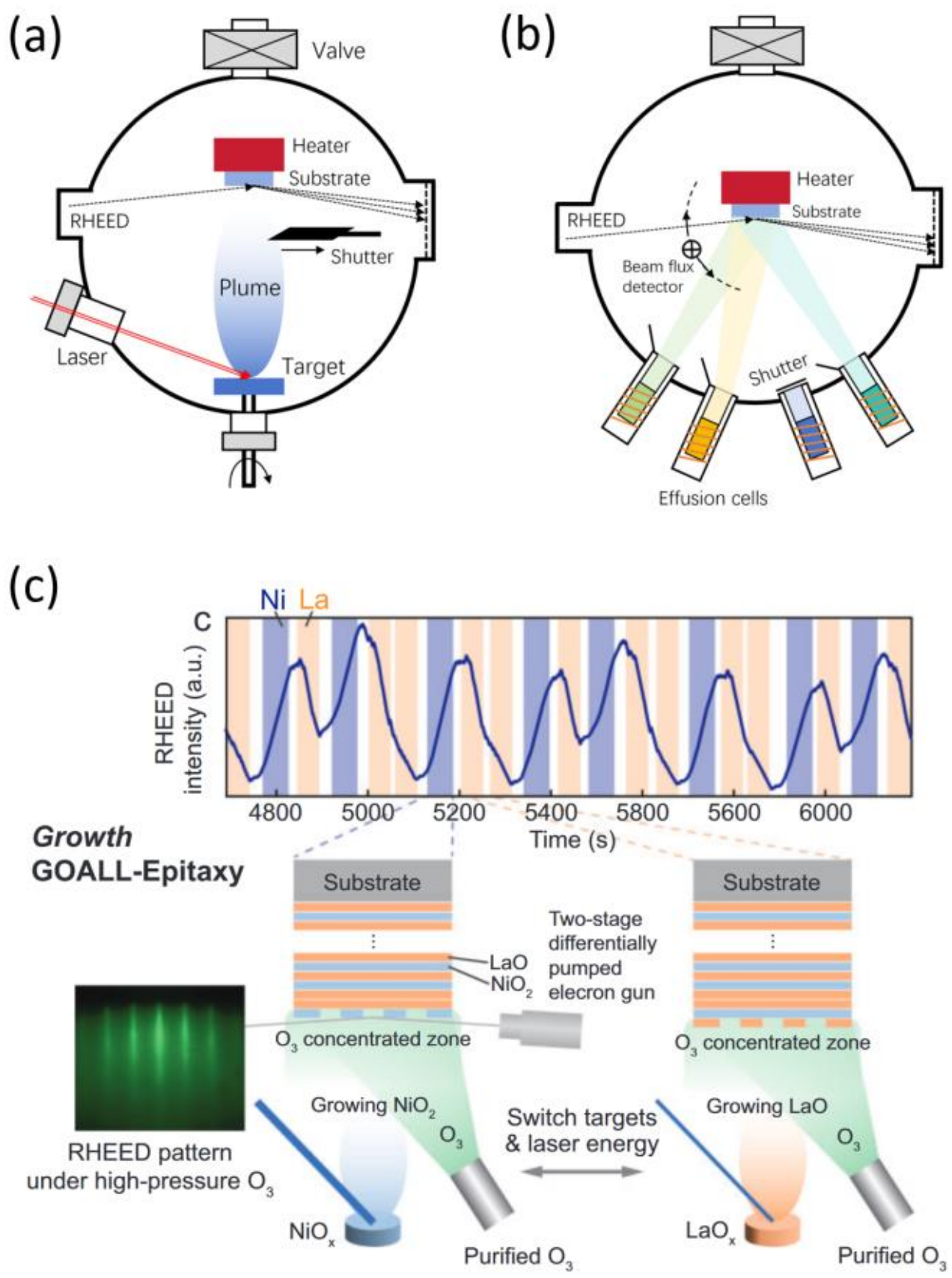

**Figure 4.** Growth techniques for epitaxial $RA_3Ni_2O_7$ thin films. (a) Pulsed laser deposition (PLD). Precise control over laser fluence, target stoichiometry, and a highly oxidizing background atmosphere is imperative for stabilizing the single-phase, superconducting RP structure. (b) Molecular beam epitaxy (MBE). (c) Gigantic-oxidative atomic-layer-by-layer epitaxy (GAE). The bottom schematic illustrates how complex RP nickelate structures are constructed on an atomic scale within a GAE deposition system. Adapted from [82]. Reprinted with permission from Oxford University Press.

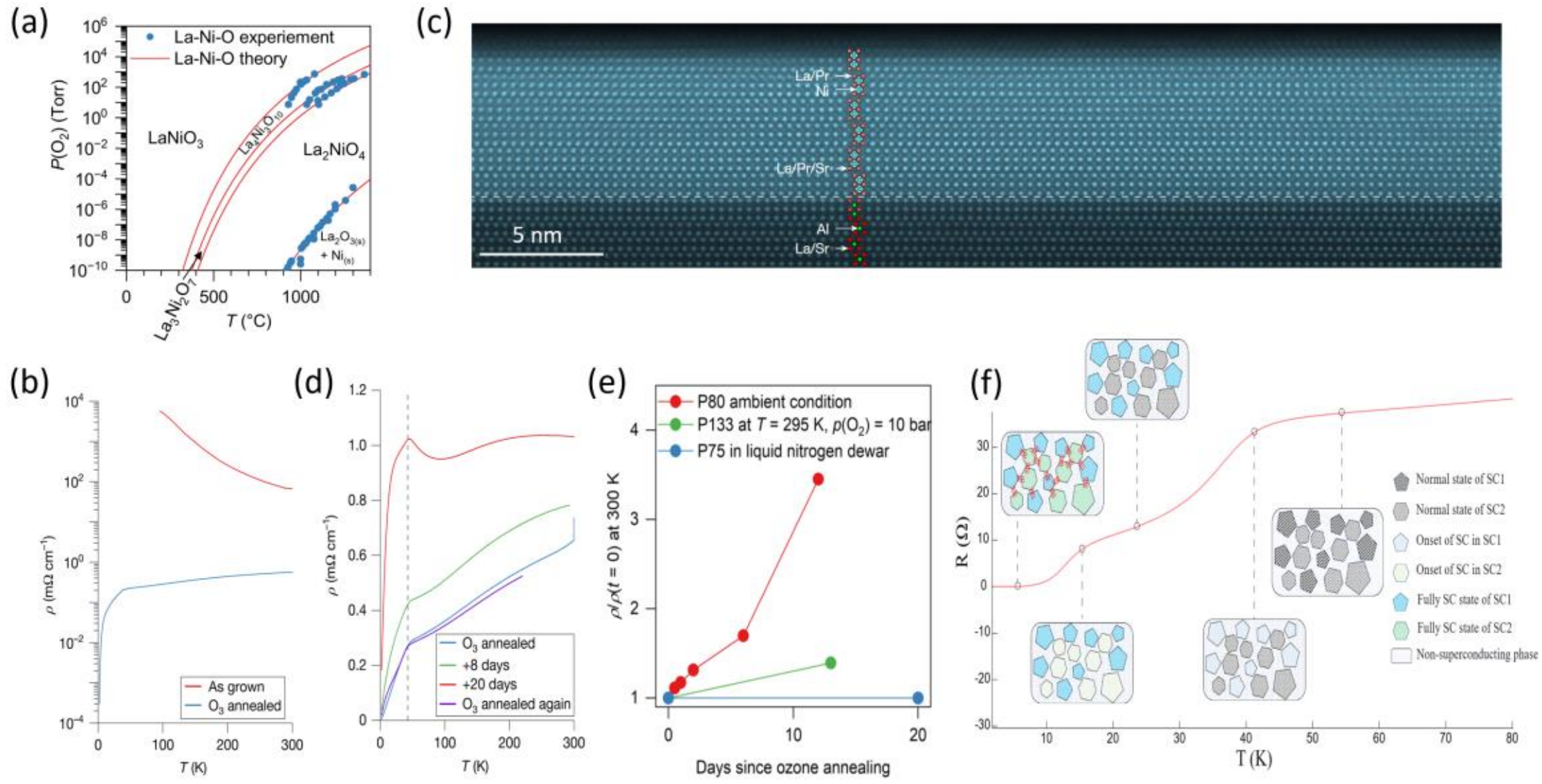

**Figure 5.** Challenges in the synthesis of $RA_3Ni_2O_7$ thin films. (a) Temperature–oxygen partial pressure phase diagram for $La_{n+1}Ni_nO_{3n+1}$, indicating a narrow thermodynamic stability window for the $n = 2$ ($La_3Ni_2O_7$) phase. (b) Necessity of strong oxidative post-annealing: $La_3Ni_2O_7$ films grown by PLD by the Hwang group are insulating as-deposited and develop superconductivity only after *ex situ* ozone annealing. (c) Suppression of parasitic RP phases by Pr substitution. High-resolution HAADF-STEM image of a $(La, Pr)_3Ni_2O_7$ film (~6.6 nm thick) over a large field of view; the dashed line marks the film–substrate interface. (d) Susceptibility of $La_3Ni_2O_7$ films to oxygen loss under ambient conditions. After 20 days of air exposure, the superconducting properties are severely degraded, although the onset $T_c$ remains relatively unchanged. (e) Cryogenic storage (e.g., immersion in liquid nitrogen) effectively suppresses the degradation of superconducting properties. (f) Two-step superconducting transition frequently observed in $RA_3Ni_2O_7$ films, attributed to granular superconductivity arising from the coexistence of two distinct superconducting phases coupled through a Josephson junction network. Red lines schematically represent Josephson junctions through which dissipationless current can tunnel. (a-e) Adapted from [4,5,25], with permission from Springer Nature. (f) From [83]. CC BY 4.0.

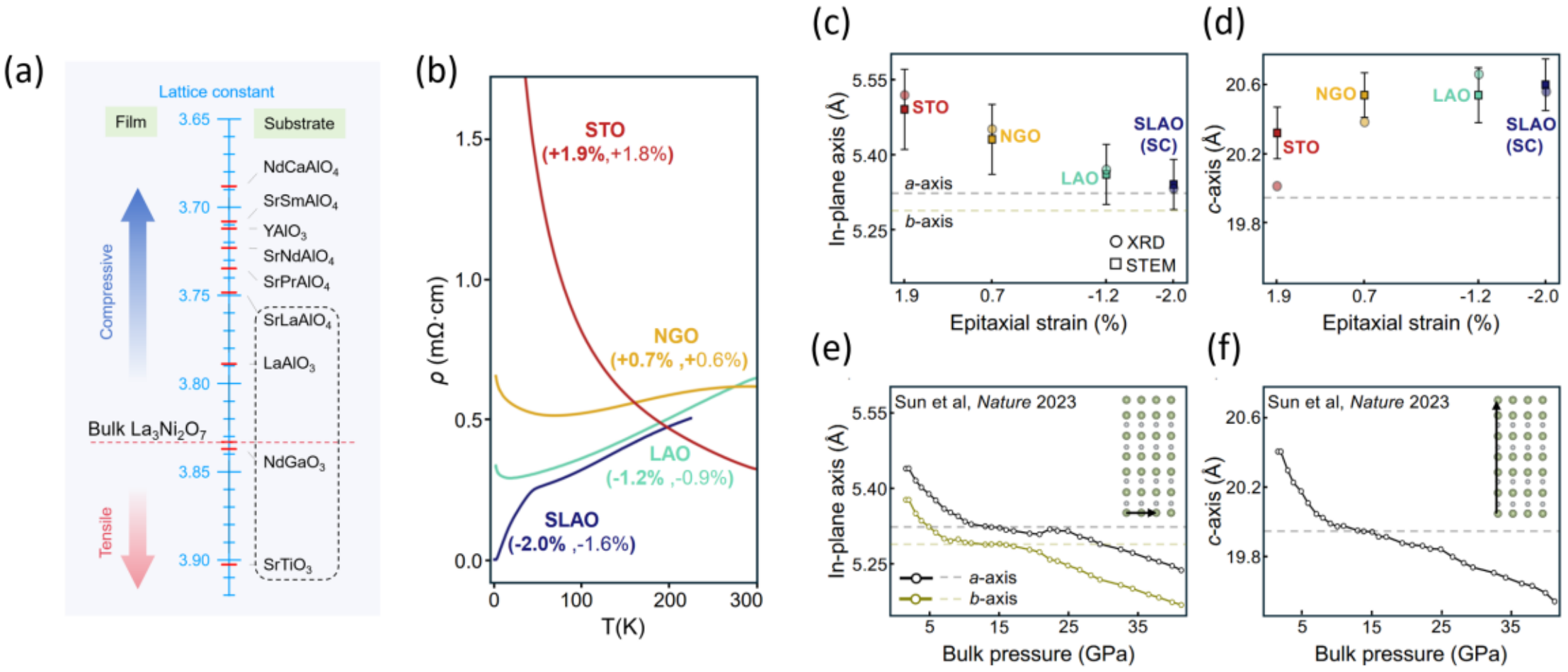

**Figure 6.** Substrate-induced lattice strain and its effect on the transport properties of $La_3Ni_2O_7$ films. (a) Lattice mismatch between $La_3Ni_2O_7$ and a range of substrates (lattice constants provided). The four substrates enclosed by black dashed boxes—$SrLaAlO_4$ (SLAO), $LaAlO_3$ (LAO), $NdGaO_3$ (NGO), and $SrTiO_3$ (STO)—have previously been used to grow $La_3Ni_2O_7$ films; SLAO and LAO impose compressive strain, whereas NGO and STO impose tensile strain. (b) Temperature-dependent resistivity of $La_3Ni_2O_7$ films epitaxially grown on the four substrates highlighted in (a); nominal and measured strain values are both indicated (negative values denote compressive strain). The onset $T_c$ for the film on SLAO is 42 K. (c) In-plane and (d) out-of-plane (c-axis) lattice constants of thin films as a function of epitaxial strain, determined by XRD (circles) and ADF-STEM (squares). Dashed lines indicate the lattice parameter values corresponding to the onset of superconductivity and structural transitions in bulk $La_3Ni_2O_7$. (e) In-plane and (f) c-axis lattice constants of bulk $La_3Ni_2O_7$ as a function of applied pressure. A comparison reveals that in-plane lattice compression is a universally required feature for the emergence of superconductivity in both bulk and thin-film forms. (b-f) Adapted from [8], with permission from Springer Nature.

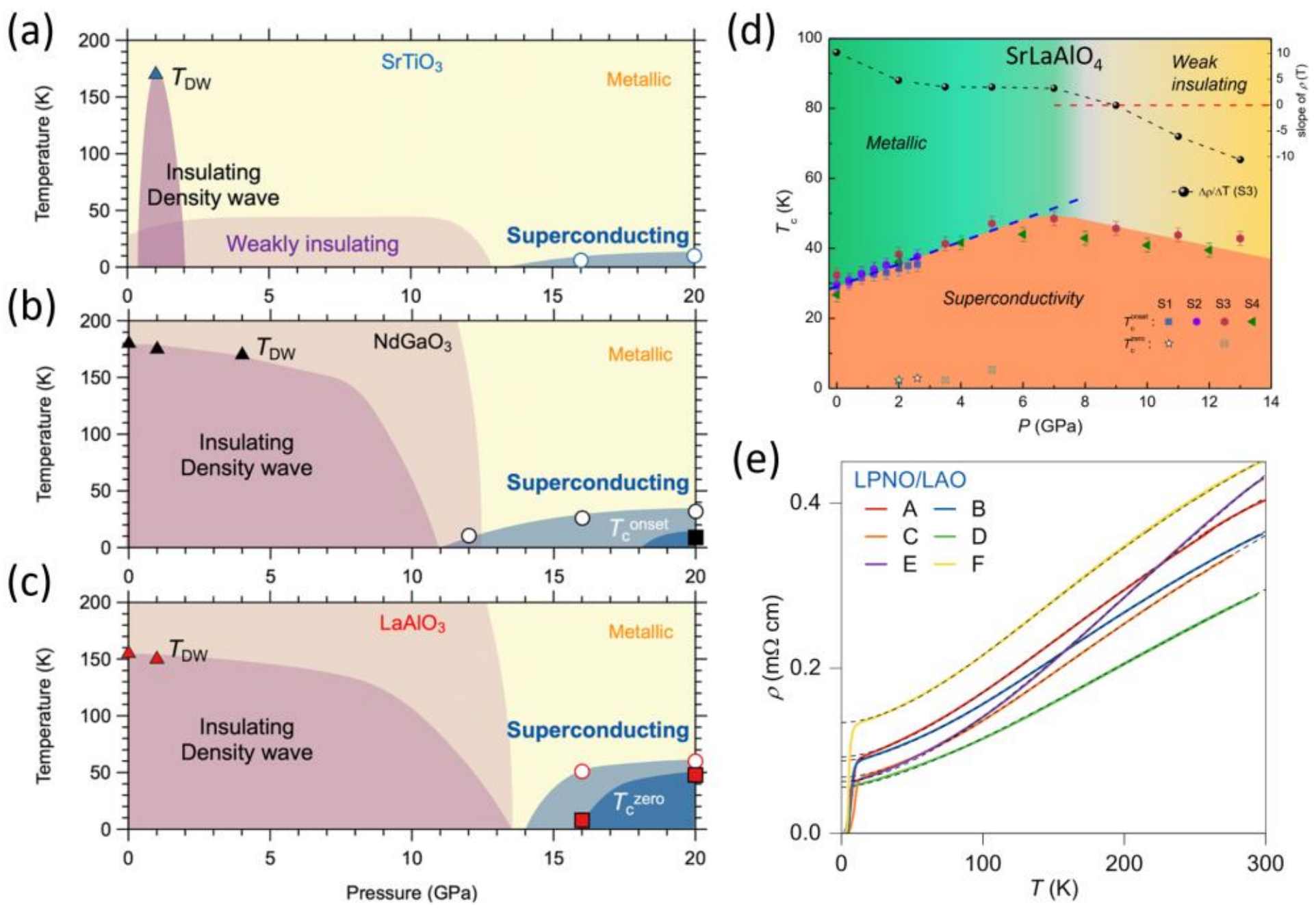

**Figure 7.** Pressure-induced superconductivity in $RA_3Ni_2O_7$ films on alternative substrates. Temperature-pressure (*T-P*) phase diagrams for $La_3Ni_2O_7$ films epitaxially grown on (a) STO, (b) NGO, and (c) LAO substrates. (d) *T-P* phase diagram for a $(La,Pr)_3Ni_2O_7$/$(La,Sr)_3Ni_2O_7$ heterostructure grown on SLAO. (e) Following rigorous optimization of the deposition and post-annealing protocols, ambient-pressure superconductivity can be stabilized in $La_3Ni_2O_7$/LAO films, displaying an onset $T_c$ of ~10 K. (a-d) Adapted from [7,24], with permission from Springer Nature. (e) From [6]. Reprinted with permission from Wiley.

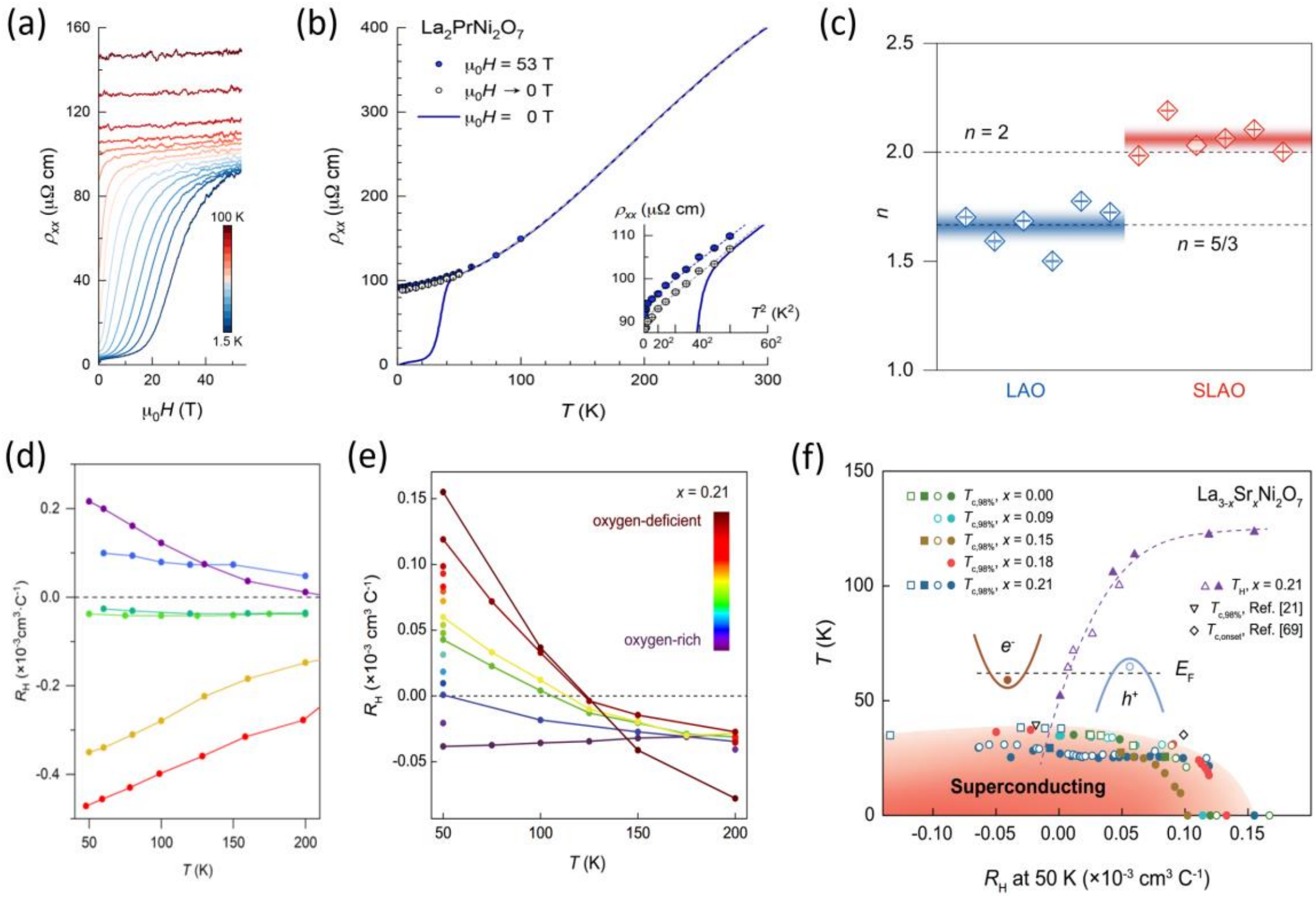

**Figure 8.** Normal-state transport properties of $RA_3Ni_2O_7$ thin films. (a) Magnetoresistance isotherms for a $La_2PrNi_2O_7$ film measured up to 53 T at various temperatures. (b) Temperature-dependent longitudinal resistivity measured at 0 T and 53 T. The gray dashed curve represents a parallel-resistor model fit ($1/\rho = 1/(\rho_0 + AT^n) + 1/\rho_{max}$) to the zero-field data between 50 and 300 K, yielding $n = 2$. Inset: $\rho_{xx}$ versus $T^2$ below 60 K. (c) Normal-state resistivity of $La_2PrNi_2O_7$ films grown on LAO substrates, where the fit yields $n = 5/3$, in contrast to the $n = 2$ behavior observed on SLAO substrates. (d) Temperature dependence of the Hall coefficient $R_H$ for $RA_3Ni_2O_7$ films reported by different research groups. (e) $R_H(T)$ of $La_{2.79}Sr_{0.21}Ni_2O_7$ films; curves from bottom to top correspond to decreasing oxygen content. (f) Evolution of $T_c$ as a function of $R_H$ (50 K) for $La_{3-x}Sr_xNi_2O_{7-\delta}$ films. Varying marker styles denote different sample batches, while colors indicate the Sr doping level. $T_H$ marks the electron-hole crossover temperature. Inset: Schematic of a two-band model illustrating electron and hole dominance on opposite sides of the $T_H$ boundary; the black dashed line indicates the Fermi level, $E_F$. (a-b) Adapted from [19], with permission from Springer Nature. (c) From [6]. Reprinted with permission from Wiley. (d) From [23]. CC BY 4.0. (e-f) Reproduced figure with permission from [22], Copyright (2026) by the American Physical Society.

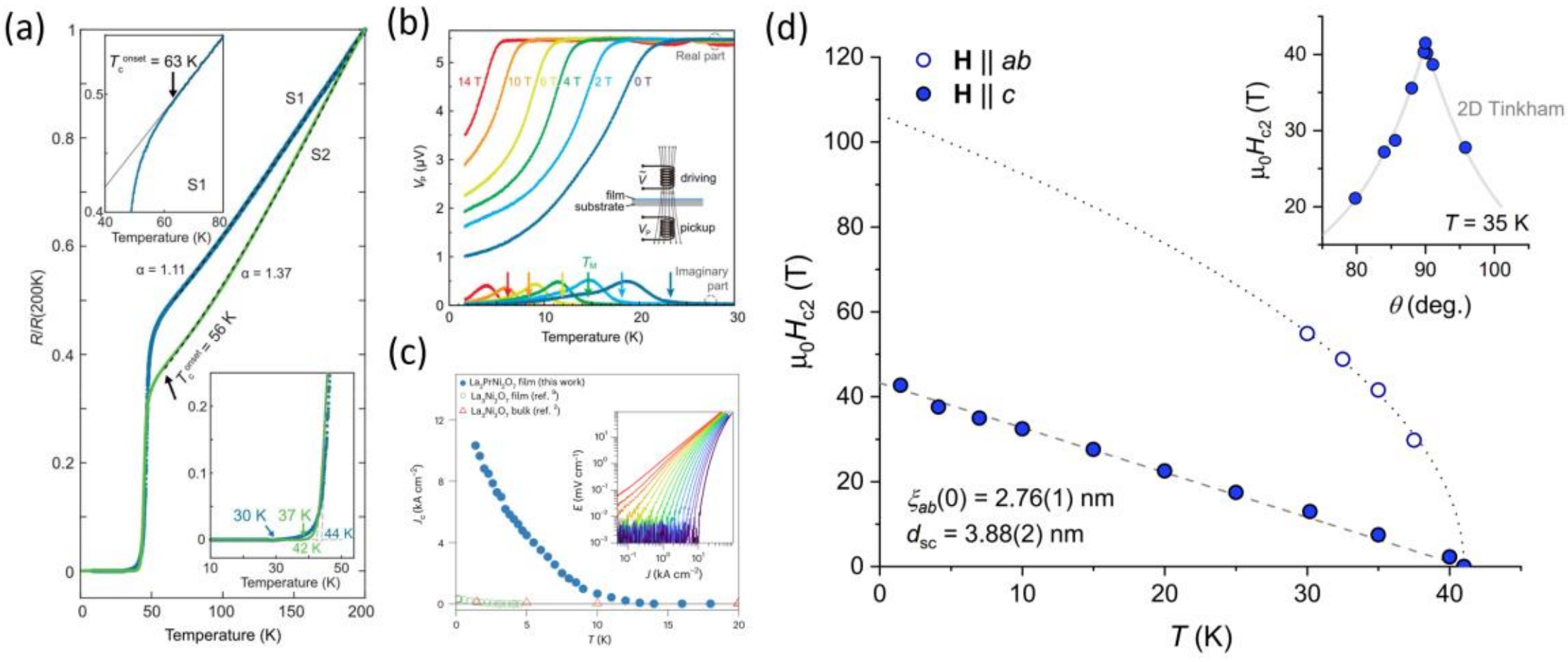

**Figure 9.** Superconducting properties of $RA_3Ni_2O_7$ thin films. (a) Temperature-dependent resistance $R(T)$ of $(La,Pr)_3Ni_2O_7$ films; the normal-state resistivity (dashed line) is fitted as $R = cT^{\alpha} + R_0$. Upper inset: Expanded view near the superconducting transition; the onset $T_c$ is 63 K, the highest value currently reported for $RA_3Ni_2O_7$ thin films. Lower inset: Expanded view of the zero-resistance regime. (b) Temperature dependence of the real (upper) and imaginary (lower) components of the mutual inductance voltage $V_P$ measured under varying applied magnetic fields. The real component reflects diamagnetic screening, and the onset of the imaginary component (arrow) defines the vortex melting temperature. (c) Critical current density $J_c$ for the $RA_3Ni_2O_7$ thin film with the highest $J_c$ reported to date; inset: log–log plot of electric field $E$ versus current density $J$. (d) Upper critical field ($\mu_0 H_{c2}$) of $La_2PrNi_2O_7$ films extracted from high-magnetic-field transport data. Filled and open symbols correspond to fields applied perpendicular and parallel to the film surface, respectively, with curves fitted using the Ginzburg–Landau formalism. Inset: Angular dependence of the critical field confirming that the magnetotransport behavior closely follows the 2D Tinkham model. (a-b) Adapted from [18]. Reprinted with permission from Oxide University Press. (c-d) Adapted from [19,25], with permission from Springer Nature.

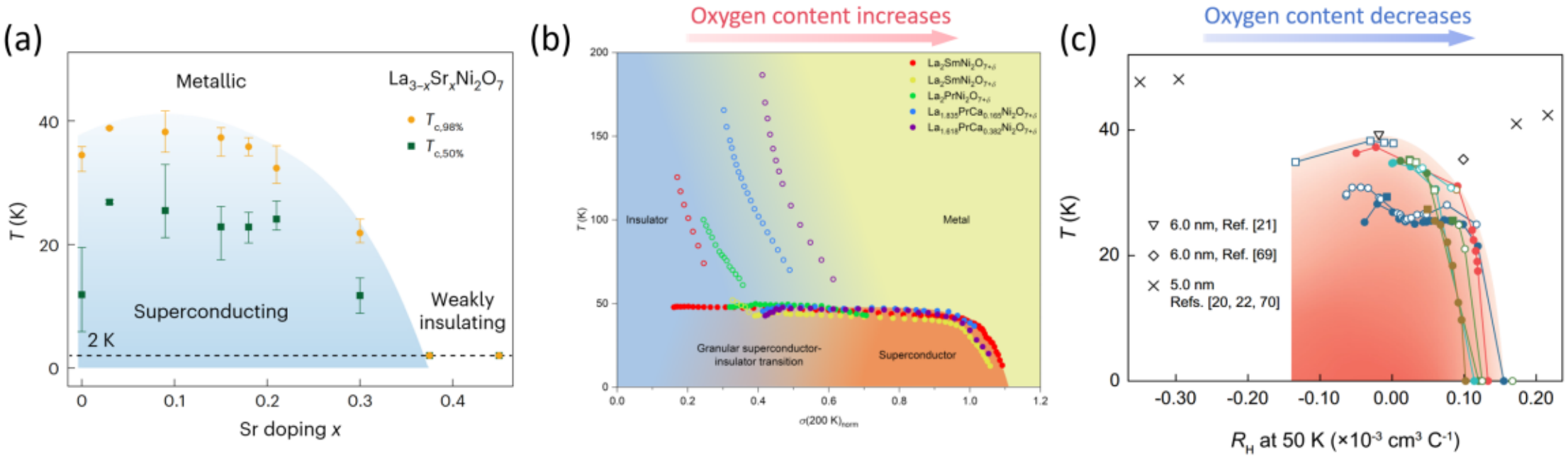

**Figure 10.** Doping-dependent superconducting phase diagrams. (a) Phase diagram of Sr-doped $La_3Ni_2O_7$ films, revealing an incomplete $T_c$ dome. The onset $T_c$ remains remarkably stable up to a doping level of $x \sim 0.21$, followed by a sharp suppression near $x = 0.38$. The optimally doped

composition exhibits a $T_c$ of ~42 K. (b) Evolution of the electronic ground state with increasing oxygen content (left to right). Progressing from a highly oxygen-deficient state towards stoichiometric oxygen concentration, the system transitions from an insulator with embedded granular superconductivity (blue) into a robust macroscopic superconducting state (red). Excessive oxygen intercalation beyond stoichiometry suppresses superconductivity, driving the system into a normal metallic state (yellow). (c) Suppression of superconductivity with decreasing oxygen content (left to right): progressive oxygen removal continuously degrades the superconducting properties of the films. (a) Adapted from [20], with permission from Springer Nature. (b) From [23]. CC BY 4.0. (c) Reproduced figure with permission from [22], Copyright (2026) by the American Physical Society.

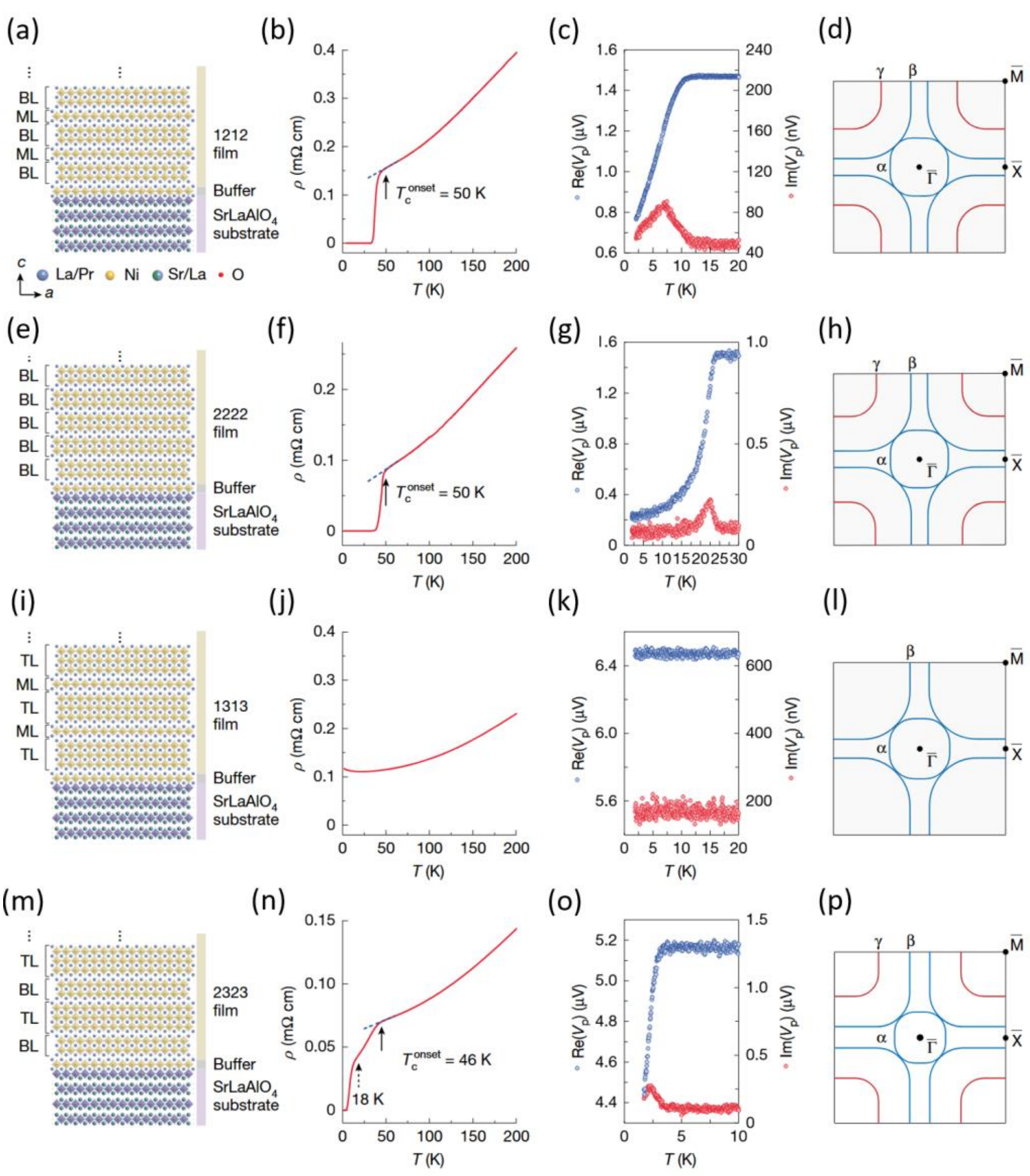

**Figure 11.** Structural and electronic comparison of 1212, 2222, 1313, and 2323 nickelate thin films.

(a, e, i, m) Schematics of the four heterostructure configurations epitaxially grown on SLAO substrates. (b, f, j, n) Electrical transport measurements for each configuration; the 1313 heterostructure does not exhibit superconductivity. (c, g, k, o) Diamagnetic response of the samples; blue and red curves represent the real and imaginary parts of the mutual inductance signal, respectively. (d, h, l, p) Fermi surface topology derived from ARPES measurements; blue lines denote the $\alpha$ and $\beta$ pockets, and red lines denote the $\gamma$ pocket. Adapted from [9], with permission from Springer Nature.

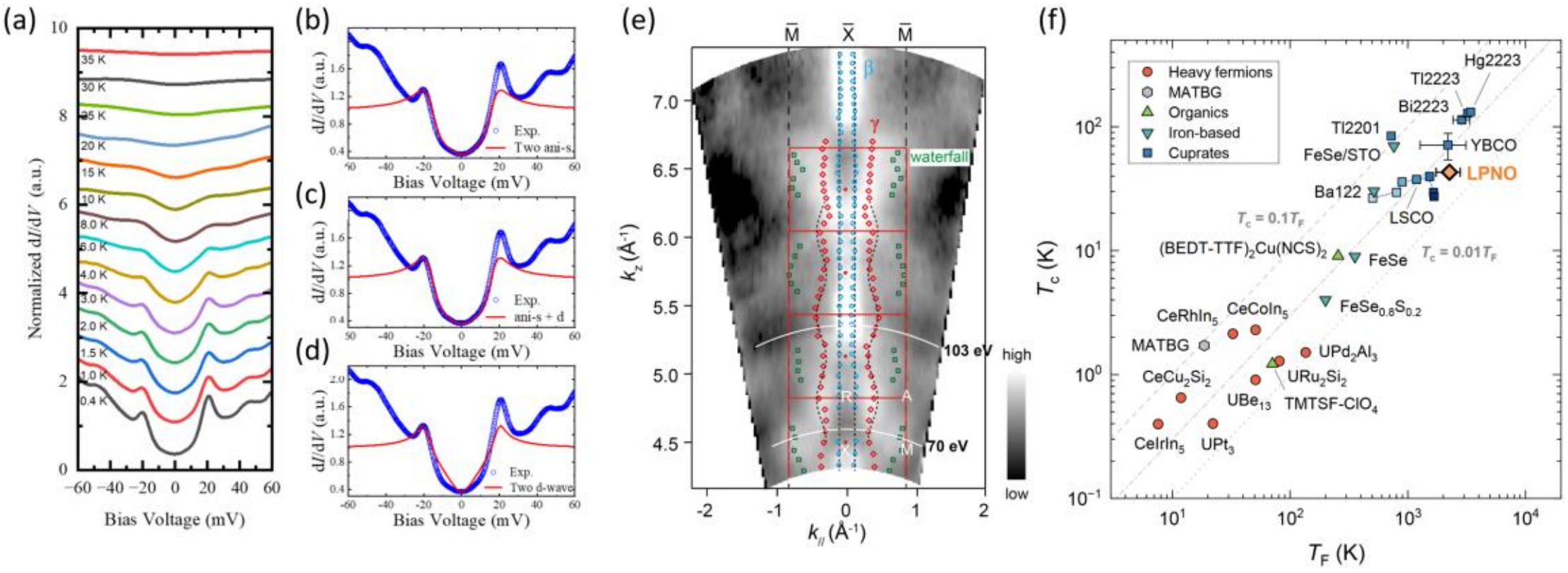

**Figure 12.** Superconducting energy gap and electronic correlations in bilayer nickelates. (a) Tunneling spectra of $La_2PrNi_2O_7$ thin films at selected temperatures. (b–d) Fits to the 0.4 K spectrum using the two-gap Dynes model: fits in (b) and (c) reproduce the measured spectrum well, whereas the fit in (d) fails to capture the principal spectral features. (e) ARPES intensity map at the Fermi energy ($E_F$) in the $k_z$-$k_{//}$ plane along the $\overline{\mathrm{M}}-\overline{\mathrm{X}}$ high-symmetry direction. The light blue, red, and green markers trace the dispersion of the $\beta$ band, $\gamma$ band, and the waterfall feature, respectively. The color scale denotes photoemission intensity. (f) Uemura plot ($T_c$ versus effective Fermi temperature, $T_F$) comparing various families of strongly correlated superconductors. The positioning of the $La_2PrNi_2O_7$ system demonstrates that superconductivity in bilayer RP nickelates falls firmly within the strongly correlated regime. (a-e) From [13,17]. CC BY 4.0. (f) Adapted from [19], with permission from Springer Nature.

## Data availability statement

All data that support the findings of this study are included within the article (and any supplementary files).

## Acknowledgement

This work was financially supported by National Natural Science Foundation of China (Grant No. 12504226).

## ORCID iDs

Meng Zhang: 0000-0001-7609-6393